# Decay Constants of Pseudoscalar Mesons

Swee Ping Chia
spchia@um.edu.my

*Physics Department, University of Malaya*
*50603 Kuala Lumpur, Malaysia*

For purely leptonic decays of pseudoscalar mesons, the decay rates are related to the product of the relevant weak interaction-based CKM matrix element of the constituent quarks on the one hand, and the strong interaction parameter, the decay constant, which is related to the overlap of the quark and antiquark in the meson on the other hand. The decay constants for these mesons can thus be estimated from the decay rates of the respective dominant decay modes. The decay constants so obtained are used to estimate the decay rates of the less dominant modes, which are in good agreement with experimentally measured values. We also predict the decay rates for τ-lepton.

## I. INTRODUCTION

The purely leptonic decays of $\pi^\pm$, $K^\pm$, $D^\pm$, $D_s^\pm$, and $B^\pm$ pseudoscalar mesons have been well measured experimentally. The measured decay rates depend on the Cabibbo-Kobayashi-Maskawa (CKM) mixing matrix element of the constituent quarks and a strong interaction parameter related to the overlap of the quark and antiquark wave-functions in the meson, called the decay constant $f_P$ [1].

A charged meson can decay to a lepton-neutrino pair via a virtual W boson, as illustrated in Fig. 1.. The decay rate for the process in Fig.1 can be straightforwardly calculated, and is given by

$$\Gamma(P \to \ell\nu) = \frac{G_F^2}{8\pi} f_P^2 m_\ell^2 M_P \left(1 - \frac{m_\ell^2}{M_P^2}\right)^2 |V_{21}|^2 \tag{1}$$

Here $M_P$ is the P meson mass, $m_\ell$ is the mass of the lepton $\ell$, $V_{21}$ is the CKM matrix element between the constituent quarks $q_1$ and $q_2$ in P, and $G_F$ is the Fermi coupling constant. The parameter $f_P$ is the decay constant which is related to the wave-function overlap of the quark and antiquark system.

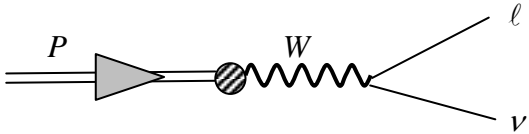

Fig. 1. Diagram representing $P \to \ell\nu$ in this model, where P is $q_1\bar{q}_2$.

## II. CALCULATION OF DECAY CONSTANTS FROM THE EXPERIMENTAL DECAY RATES

The decay rate for the process $P \to \ell \nu$ can be inferred from the lifetime of $P$ and the relevant decay branching ratio [2]. Table 1 gives the leptonic decay rates for the prominent decay modes considered.

TABLE 1. Decay rates for the prominent leptonic decays of pseudoscalar mesons.

| Particle | Mean Life (s) | Decay Modes | Branching Ratio | Decay Rate (MeV) |
|---|---|---|---|---|
| $\pi^+$ | $2.6033(5) \times 10^{-8}$ | $\mu^+ \nu_\mu$ | 99.98770(4) % | $2.528064(10) \times 10^{-14}$ |
| $K^+$ | $1.2380(21) \times 10^{-8}$ | $\mu^+ \nu_\mu$ | 63.55(11) % | $3.378(6) \times 10^{-14}$ |
| $D^+$ | $1.040(7) \times 10^{-12}$ | $\mu^+ \nu_\mu$ | $3.82(33) \times 10^{-4}$ | $2.42(21) \times 10^{-13}$ |
| $D_s^+$ | $0.500(7) \times 10^{-12}$ | $\mu^+ \nu_\mu$ | $5.90(33) \times 10^{-3}$ | $7.8(4) \times 10^{-12}$ |
| $B^+$ | $1.641(8) \times 10^{-12}$ | $\tau^+ \nu_\tau$ | 1.65(34) % | $6.6(14) \times 10^{-14}$ |

To obtain the decay constant $f_P$, we need information on the CKM matrix elements as accurately as possible. The following values are obtained from Particle Data Group's 2012 edition [3]:

$$|\mathbb{V}| = \begin{pmatrix} V_{ud} & V_{us} & V_{ub} \\ V_{cd} & V_{cs} & V_{cb} \\ V_{td} & V_{ts} & V_{tb} \end{pmatrix} = \begin{pmatrix} 0.97425(22) & 0.2252(9) & 0.00415(49) \\ 0.230(11) & 0.9723(26) & 0.0409(11) \\ 0.0084(6) & 0.0429(26) & 0.99904(11) \end{pmatrix} \quad (2)$$

The elements $V_{cs}$ and $V_{tb}$ are not well measured. Instead, they are calculated by exploiting the unitary relations.

The decay constant $f_P$ is then calculated from Eq. (1). For the dominant decay modes of pseudoscalar mesons, the decay constant values calculated are listed in Table 2.

TABLE 2. Values of decay constants $f_P$ estimated from leptonic decay rates of the dominant decay modes.

| Decay Modes | Decay Rate (MeV) | Decay Constant $f_P$ |
|---|---|---|
| $\pi^+ \to \mu^+ \nu_\mu$ | $2.528064(10) \times 10^{-14}$ | 133.008 (30) |
| $K^+ \to \mu^+ \nu_\mu$ | $3.378(6) \times 10^{-14}$ | 156.6 (8) |
| $D^+ \to \mu^+ \nu_\mu$ | $2.42(21) \times 10^{-13}$ | 201(18) |
| $D_s^+ \to \mu^+ \nu_\mu$ | $7.8(4) \times 10^{-12}$ | 264(7) |
| $B^+ \to \tau^+ \nu_\tau$ | $6.6(14) \times 10^{-14}$ | 232(52) |

The values for the decay constants obtained above are in agreement with earlier experimental results [1, 4-10] and various theoretical calculations [11-17]. The large uncertainty in $B^+ \to \tau^+ \nu_\tau$ is because of the relatively larger experimental error in $V_{ub}$.

## III. COMPARISON OF CALCULATED RATES WITH MEASURED RATES FOR THE LESS DOMINANT MODES

Using these values of the decay constants, we can calculate the decay rates for the less dominant decay modes of the corresponding decays. Table 3 shows the calculated rates against the measured rates for these less dominant modes [2].

TABLE 3. Calculated decay rates against measured rates for the less dominant decay modes.

| Decay Modes | Calculated Rate (MeV) | Measured Rate (MeV) |
|---|---|---|
| $\pi^+ \to e^+ \nu_e$ | $3.3125(30) \times 10^{-18}$ | $3.110(10) \times 10^{-18}$ |
| $K^+ \to e^+ \nu_e$ | $8.68(16) \times 10^{-19}$ | $8.41(4) \times 10^{-19}$ |
| $D^+ \to e^+ \nu_e$ | $5.65(16) \times 10^{-18}$ | $< 5.6 \times 10^{-15}$ |
| $D^+ \to \tau^+ \nu_\tau$ | $6.40(18) \times 10^{-13}$ | $< 7.6 \times 10^{-13}$ |
| $D_s^+ \to e^+ \nu_e$ | $1.83(11) \times 10^{-16}$ | $< 1.6 \times 10^{-13}$ |
| $D_s^+ \to \tau^+ \nu_\tau$ | $7.6(4) \times 10^{-11}$ | $7.1(4) \times 10^{-11}$ |
| $B^+ \to e^+ \nu_e$ | $7(5) \times 10^{-21}$ | $< 3.9 \times 10^{-16}$ |
| $B^+ \to \mu^+ \nu_\mu$ | $3.0(20) \times 10^{-16}$ | $< 4.0 \times 10^{-16}$ |

The agreement is excellent. It is noted that the calculated values are very close to the experimental upper bounds for the following decay modes: $D^+ \to \tau^+ \nu_\tau$ and $B^+ \to \mu^+ \nu_\mu$.

## IV. EXTENSION TO τ-LEPTON DECAYS

Eq. (1) can be extended to describe τ-lepton decays:

$$\Gamma(\tau \to P^- \nu_\tau) = \frac{G_F^2}{16\pi} f_P^2 m_\tau^3 M_P \left(1 - \frac{M_P^2}{m_\tau^2}\right)^2 |V_{21}|^2 \qquad (3)$$

Using the same set of values for the decay constants obtained in Table 2, the τ-lepton decay rates calculated are displayed in Table 4 against the measured rates [2]. The agreement is good.

TABLE 4. Calculated decay rates of $\tau$-lepton against measured rates.

| Decay Modes | Calculated Rate (MeV) | Measured Rate (MeV) |
|---|---|---|
| $\tau^- \to \pi^- \nu_\tau$ | $2.5180(23) \times 10^{-10}$ | $2.4530(14) \times 10^{-10}$ |
| $\tau^- \to K^- \nu_\tau$ | $1.608(29) \times 10^{-11}$ | $1.385(23) \times 10^{-11}$ |

## V. CONCLUSION

We have presented here a straightforward way to estimate the decay constants for pseudoscalar mesons. The results are consistent with earlier experimental estimates and theoretical calculations. Values of the decay constants are used to calculate the decay rates for the less dominant decay modes, yielding good agreement with experimental values. The same decay constants are also utilized to generate the decay rates of τ-lepton.

## ACKNOWLEDGMENTS

This research was carried out under the Research Grant UMRG049/09AFR from University of Malaya. Part of this research was carried while on research leave at Academia Sinica, Taiwan. The author wishes to thank Prof. Hai-Yang Cheng for valuable discussion.